\documentclass[useAMS,usenatbib]{mn2e}

\usepackage[breaklinks,colorlinks,citecolor=blue,linkcolor=red]{hyperref}

\usepackage{graphicx}
\usepackage{amsmath}
\usepackage{amssymb}
\usepackage{color}
\usepackage{mathtools}
\usepackage{ulem}

\usepackage[all]{hypcap}

\newcommand\msun{\, \rm M_\odot}

\newcommand\be{\begin{equation}}
\newcommand\ee{\end{equation}}

\title[Constraining NS radii in BH-NS star mergers]{Constraining neutron star radii in black hole-neutron star mergers from their electromagnetic counterparts}
\author[G. Fragione \& A. Loeb]{\parbox{\textwidth}{Giacomo Fragione$^{1,2}$\thanks{E-mail: giacomo.fragione@northwestern.edu}, Abraham Loeb$^{3}$}\\
\ \\
$^1$Department of Physics \& Astronomy, Northwestern University, Evanston, IL 60202, USA\\
$^2$Center for Interdisciplinary Exploration \& Research in Astrophysics (CIERA), Evanston, IL 60202, USA\\
$^3$Astronomy Department, Harvard University, 60 Garden St., Cambridge, MA 02138, USA}

\begin{document}

\maketitle

\begin{abstract}
Mergers of black hole (BH) and neutron star (NS) binaries are of interest since the emission of gravitational waves (GWs) can be followed by an electromagnetic (EM) counterpart, which could power short gamma-ray bursts. Until now, LIGO/Virgo has only observed a candidate BH-NS event, GW190426\_152155, which was not followed by any EM counterpart. We discuss how the presence (absence) of a remnant disk, which powers the EM counterpart, can be used along with spin measurements by LIGO/Virgo to derive a lower (upper) limit on the radius of the NS. For the case of GW190426\_152155, large measurement errors on the spin and mass ratio prevent from placing an upper limit on the NS radius. Our proposed method works best when the aligned component of the BH spin (with respect to the orbital angular momentum) is the largest, and can be used to complement the information that can be extracted from the GW signal to derive valuable information on the NS equation of state.
\end{abstract}


\section{Introduction}
\label{sect:intro}

Together with GWTC-1, from the first two observational runs \citep{lvc2019cat}, the new candidate events presented in \citet{AbbottAbbott2020a}, from the first half of the third observational run, comprise GWTC-2. Among its events, there are both black holes (BHs) and neutron stars (NSs) merging in binary systems. Thanks to the growing number of detected events, compact objects can be constrained with unparalleled precision and gravitational wave (GW) events provide an unprecedented opportunity to probe fundamental physics \citep{AbbottAbbott2020b,AbbottAbbott2020c}.

The origin of binary mergers is still highly debated. Several possible scenarios could potentially account for most of the observed events  \citep[e.g.,][]{antoper12,belcz2016,askar17,bart17,ll18,baner18,fragk2018,rod18,fragg2019,fragk2019,ham2019,krem2019,rasskoc2019}. Since several models account for roughly the same rate, first analyses of the LIGO/Virgo data have shown that the observed population is likely composed of mergers originated in more than one scenario \citep[e.g.,][]{WongBreivik2020,ZevinBavera2020}. The contribution of different astrophysical channels will be hopefully disentangled using a combination of the mass, spin, redshift, and eccentricity distributions in the upcoming years.

Despite the growing number of events, there is only a candidate BH-NS merger, namely GW190426\_152155. Therefore, LIGO has only set a $90\%$ upper limit of $\sim 600$ Gpc$^{-3}$ yr$^{-1}$ on the rate of BH-NS mergers. As for BH-BH and NS-NS systems, the origin of BH-NS binaries is still highly uncertain. While BH-NS binaries can be produced in isolation as a result of binary evolution \citep[e.g.,][]{demink2016,kruc2018}, more challenging is the process of forming BH-NS binaries through dynamical assembly in star clusters. A number of authors showed that NSs are generally prevented from forming NS-NS and BH-NS binaries in a star cluster as a result of the strong heating due to BH in the cluster core \citep{frag2018,ye2019,FragioneBanerjee2020,YeFong2020}, although some authors have claimed higher rates \citep{RastelloMapelli2020,SantoliquidoMapelli2020}. Only after most of the BHs have been ejected from the cluster core, NSs can efficiently segregate in the innermost regions and possibly form binaries, that eventually merge. Recently, \citet{frl2019,FragioneLoeb2019} have shown that BH-NS mergers can be a natural outcome of the dynamical evolution of massive triple stellar systems. 

What makes BH-NS mergers interesting is the possibility that they can produce an electromagnetic (EM) counterpart after merger. The lifetime of merging BH-NS binaries spans three phases \citep[for a recent short review see][]{Foucart2020}, which include the inspiral ($\gtrsim 10^6$\,yr) due to GW emission, the merger phase ($\sim 1$\,ms) that can result in either the tidal disruption of the NS or its plunge into the BH, and, for disrupting systems only, a post-merger phase ($\sim 1$\,s) during which more matter is ejected or accreted onto the BH.

The merger will be followed by an EM counterpart only if the NS is disrupted by the tidal field of the BH, leading to mass ejection and the formation of an accretion torus around it. Otherwise, the NS plunges into the BH and the merger resembles a BH-BH merger. In this Letter, we discuss how this information can be used along with spin measurements by LIGO/Virgo to constrain the properties of BH-NS mergers. In particular, we discuss how to derive a lower (upper) limit based on the presence (absence) of an EM counterpart.

The Letter is organized as follows. In Section~\ref{sect:remmass}, we describe the prescriptions we use to compute whether a BH-NS merger produces an accretion disk. In Section~\ref{sect:results}, we discuss how to use this information to constrain the properties of merging BHs and NSs, and apply our method to the candidate event GW190426\_152155. Finally, we draw our conclusions, in Section~\ref{sect:concl}.

\section{Remnant mass in black hole-neutron star mergers}
\label{sect:remmass}

\begin{figure} 
\centering
\includegraphics[scale=0.565]{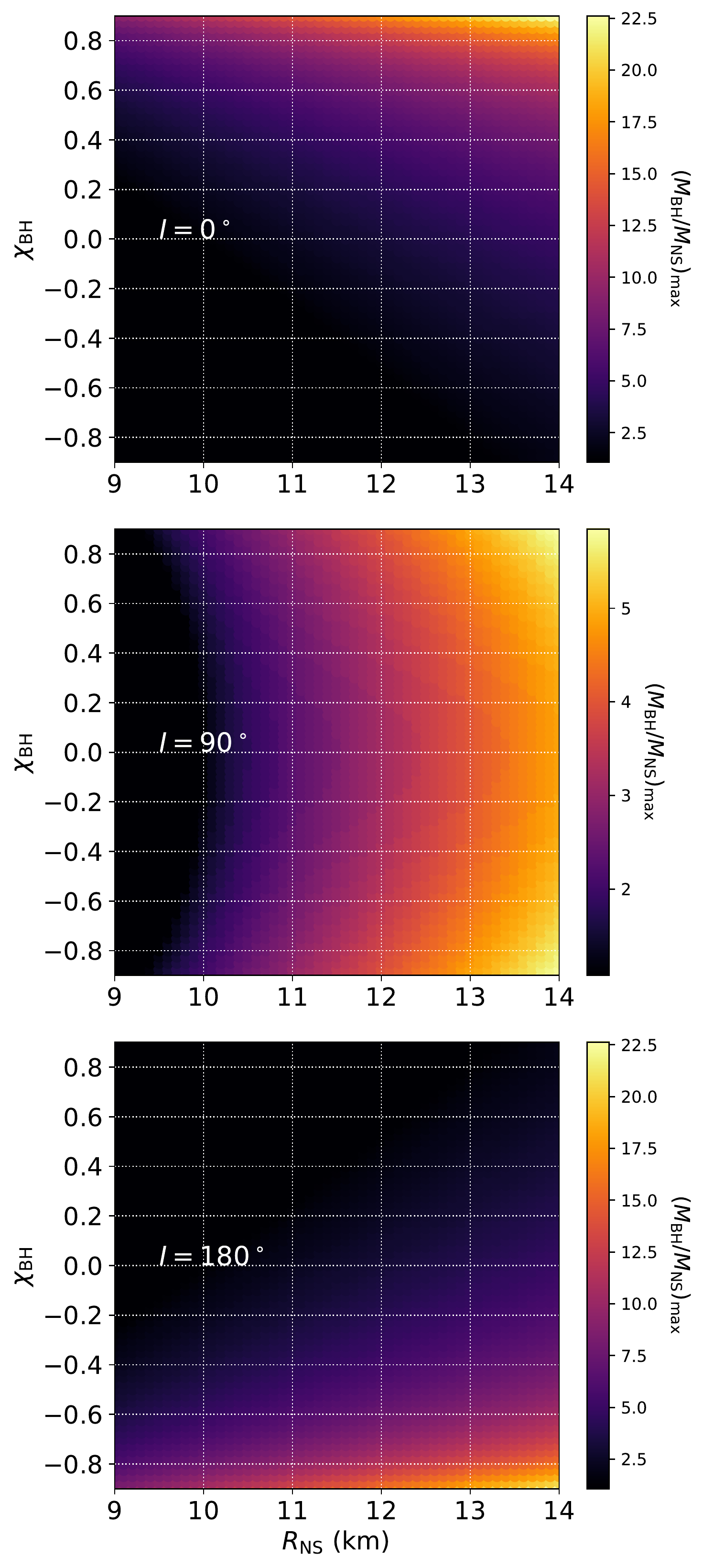}
\caption{Maximum value of the mass ratio $M_{\rm BH}/M_{\rm NS}$ for which a BH-NS system will disrupt as a function of the NS radius ($R_{\rm NS}$) and BH spin ($\chi_{\rm BH}$), assuming $M_{\rm NS}=1.3\msun$, for different inclinations $I$ between the BH spin and the orbital angular momentum. Top: $I=0^\circ$; center: $I=90^\circ$; bottom: $I=180^\circ$.}
\label{fig:plot1}
\end{figure}

\begin{figure*} 
\centering
\includegraphics[scale=0.625]{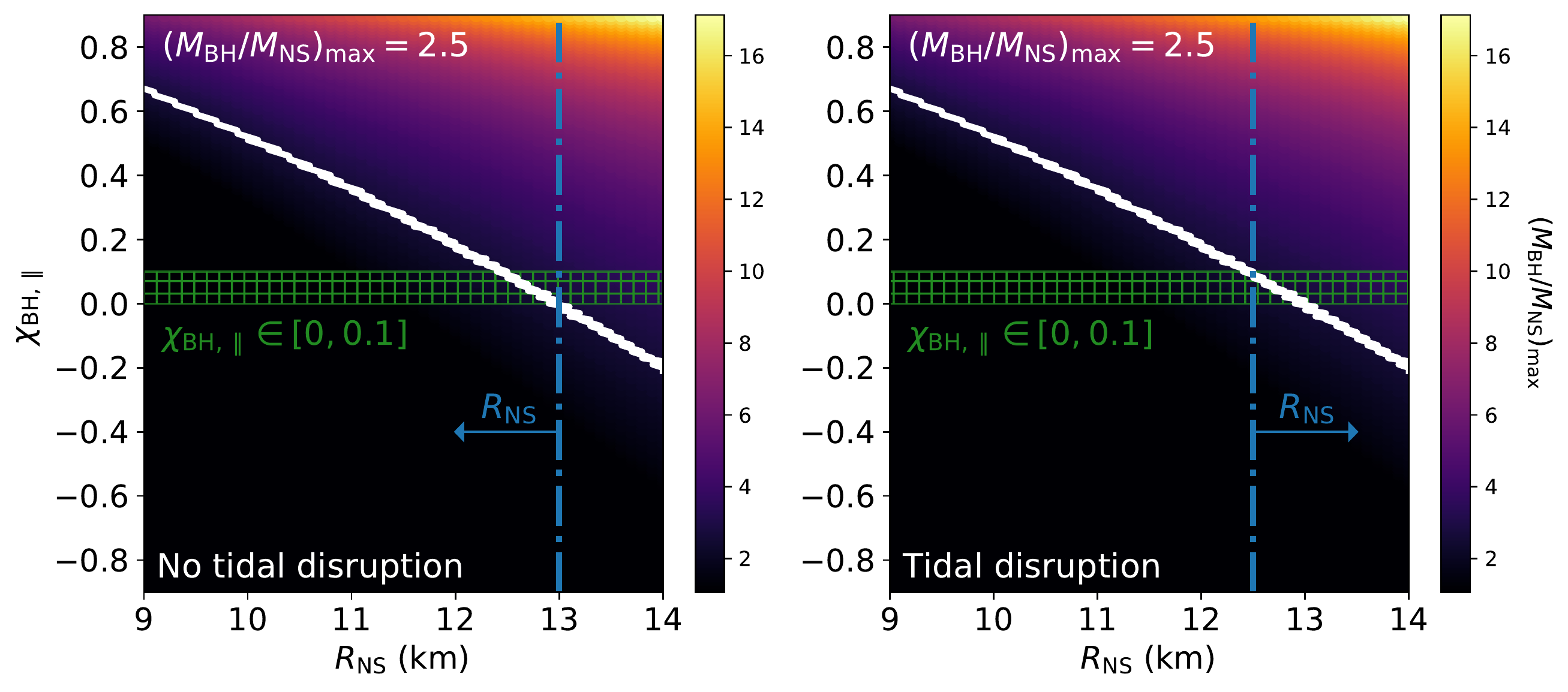}
\caption{Proof-of-concept case on how to constrain the NS radius by using the information on the effective spin and the maximum value of the mass ratio $M_{\rm BH}/M_{\rm NS}$ for which a BH-NS system disrupts the NS. We consider $M_{\rm NS}=1.5\msun$ and assume $M_{\rm BH}/M_{\rm NS}=2.5$ and that the aligned component of the BH spin is measured to $\chi_{\rm BH,\parallel}\in[0,0.1]$. Left panel: no observation of tidal disruption results in an upper limit on the NS radius; right panel: tidal disruption of the NS results in a lower limit on the NS radius.}
\label{fig:plot2}
\end{figure*}

To compute whether a BH-NS merger produces an EM signature, we compute the remnant baryon mass ($M_{\rm rem}$) outside the BH after merger. If $M_{\rm rem}>0$, a disk is formed and there is EM emission after merger \citep{Foucart2012,FoucartDeaton2013}. We use the remnant mass\footnote{In units of the initial mass of the NS.} as calibrated by \citet{FoucartHinderer2018} using numerical relativity calculations\footnote{Assuming circular binaries. For examples of general relativistic simulations of eccentric BH-NS interactions, see \citet{StephensEast2011}.},
\begin{equation}
\hat{M}_{\rm rem}=\left[\max\left(\alpha \frac{1-2C_{\rm NS}}{\eta^{1/3}}-\beta\hat{R}_{\rm ISCO}\frac{C_{\rm NS}}{\eta}+\gamma,0\right)\right]^\delta\,,
\label{eqn:mrem}
\end{equation}
where $C_{\rm NS}=GM_{\rm NS}/(R_{\rm NS}c^2)$ is the NS compaction, which depends on the equation of state,
\begin{equation}
\eta=\frac{Q}{(1+Q)^2}\,,
\end{equation}
with $Q=M_{\rm BH}/M_{\rm NS}$, is the symmetric mass ratio, and (in units $G=c=1$),
\begin{equation}
\hat{R}_{\rm ISCO}=\frac{R_{\rm ISCO}}{M_{\rm BH}}=3+Z_2-{\rm sgn}(\chi_{\rm BH})\sqrt{(3-Z_1)(3+Z_1+3Z_2)}\,,
\end{equation}
is the innermost stable circular orbit (ISCO) radius \citep{BardeenPress1972} with 
\begin{equation}
Z_1=1+(1-\chi_{\rm BH}^2)^{1/3}[(1+\chi_{\rm BH})^{1/3}+(1-\chi_{\rm BH})^{1/3}]
\end{equation}
\begin{equation}
Z_2=\sqrt{3\chi_{\rm BH}^2+Z_1^2}\,,
\end{equation}
as a solution of
\begin{equation}
Z(r)=(r(r-6))^2-a^2(2r(3r+14)-9a^2)=0\,.
\end{equation}  
In Eq.~\ref{eqn:mrem}, $(\alpha,\beta,\gamma,\delta)=(0.406,0.139,0.255,1.761)$ are fitting parameters to numerical relativity simulations \citep{Foucart2012,FoucartDeaton2013,FoucartHinderer2018}. Eq.~\ref{eqn:mrem} assumes that the BH spin is aligned to the orbital angular momentum. In the case it is inclined by an angle $I$, the same fitting formulae can be used by replacing the ISCO radius by the radius of the innermost stable spherical orbit (ISSO), which is the solution of \citep{StoneLoeb2013}
\begin{equation}
S(r)=r^8Z(r)+a^2(1-C^2)(a^2(1-C^2)Y(r)-2r^4 X(r))\,,
\label{eqn:ISSO}
\end{equation}
where $C=\cos I$, and
\begin{align}
X(r)=&a^2(a^2(3a^2+4r(2r-3)) \\
&+r^2(15r(r-4)+28))-6r^4(r^2-4)\notag  \\ 
Y(r)=&a^4(a^4+r^2(7r(3r-4)+36))+6r(r-2)\\
&\times(a^6+2r^3(a^2(3r+2)+3r^2(r-2))) \notag\,.
\end{align}
Alternatively, Eq.~\ref{eqn:mrem} can be used considering only the aligned component of the BH spin
\begin{equation}
\chi_{\rm BH,\parallel}=\chi_{\rm BH}\cos \theta_{\rm BH}\,.
\end{equation}

Figure~\ref{fig:plot1} shows the maximum value of the mass ratio $M_{\rm BH}/M_{\rm NS}$ for which a BH-NS system will disrupt as a function of the NS radius ($R_{\rm NS}$) and BH spin ($\chi_{\rm BH}$), assuming $M_{\rm NS}=1.3\msun$, for different inclinations $I$ between the BH spin and the orbital angular momentum. For $I=0^\circ$, the maximum mass ratio for disrupting systems is as high as $\approx 22$ for highly-spinning BHs. In the case $I=90^\circ$, the maximum mass ratio reduces to $\approx 6$. Obviously, $I=180^\circ$ is symmetric with respect to the the case $I=0^\circ$. This is because the case $I=0^\circ$ can be used for any inclination $I$ with the substitution $\chi_{\rm BH}\xrightarrow{}\chi_{\rm BH,\parallel}$. Thus, whether or not a BH-NS merger is followed by an EM counterpart is essentially determined by the symmetric mass ratio, the NS compaction, and the aligned component of the BH spin.

\section{Constraining neutron star radius}
\label{sect:results}

To constrain the NS radius, information on the BH spin is needed. GW measurements are especially sensitive to the effective spin, which is the following combination of the BH and NS spins \citep{AbbottAbbott2016,VitaleLynch2017},
\begin{equation}
\chi_{\rm eff}=\frac{M_{\rm BH}\chi_{\rm BH}\cos \theta_{\rm BH}+M_{\rm NS}\chi_{\rm NS}\cos \theta_{\rm NS}}{M_{\rm BH}+M_{\rm NS}}\,,
\label{eqn:chieff}
\end{equation}
where
\begin{equation}
\chi_i=\frac{cS_i}{GM_i^2}
\end{equation}
is the dimensionless Kerr parameter and $S_i=|\mathbf{S_i}|$ is the magnitude of the intrinsic spin. Approximating the NS as a rotating sphere, it can be shown that
\begin{equation}
\chi_{\rm NS}\approx \left( \frac{\nu}{1\,{\rm ms}^{-1}}\right)\,,
\end{equation}
where $\nu$ is the rotational frequency of the NS. Therefore, the NS spin will be typically small, except in the case the NS is a (rare) millisecond pulsar. In the more typical case the NS is not a millisecond pulsar, and considering that $M_{\rm BH}>M_{\rm NS}$, inverting Eq.~\ref{eqn:chieff} yields
\begin{equation}
\chi_{\rm BH,\parallel}\approx\chi_{\rm eff}\frac{M_{\rm BH}+M_{\rm NS}}{M_{\rm BH}}\,.
\label{eqn:chipareff}
\end{equation}
This implies that the aligned component of the BH spin can be approximated by Eq.~\ref{eqn:chipareff} using the values inferred by LIGO/Virgo analysis pipeline whenever the NS is not a fast rotator. This information, along with the presence or not of an EM counterpart, can be used to constrain the NS radius.

Figure~\ref{fig:plot2} illustrates a proof-of-concept case on how to constrain the NS radius by using the information on the effective spin and the maximum value of the mass ratio $M_{\rm BH}/M_{\rm NS}$ for which a BH-NS system disrupts. In our example, we consider $M_{\rm NS}=1.5\msun$ and assume that the mass ratio between the merging BH and NS is $M_{\rm BH}/M_{\rm NS}=2.5$. We note that we are not taking into account the effect of the not-negligible uncertainty in the component masses, which would cause a thickening of the $M_{\rm BH}/M_{\rm NS}$ curve on the plot. Moreover, we assume that the aligned component of the BH spin is measured in the range $\chi_{\rm BH,\parallel}\in[0,0.1]$. In case the NS plunges onto the BH and no EM counterpart is observed following the merger, the allowed portion of the parameter space is below the curve that represents $(M_{\rm BH}/M_{\rm NS})_{\rm max}=2.5$ (white line). This translates into an upper limit on the NS radius. On the other hand, if the EM counterpart is observed following the BH-NS merger, the allowed portion of the parameter space is above the curve that represents $(M_{\rm BH}/M_{\rm NS})_{\rm max}=2.5$, translating into a lower limit on the NS radius. For reference, recent measurements from LIGO/Virgo and NICER constrain the NS radius to $\sim 10.5$--$14.5$\,km \citep{AbbottAbbott2018,MillerLamb2019,RaaijmakersGreif2020}.

From our proof-of-concept example, it is clear that our method works best when the aligned component of the BH spin is the largest. Obviously, the smaller the measurement errors on the spin are, the better works our method.

\subsection{Application to GW190426\_152155}

GW190426\_152155 is the candidate event with the highest false alarm rate ($1.4$\,yr$^{-1}$) among the LIGO/Virgo events in GWTC-2. Assuming it is a signal of astrophysical origin, the LIGO/Virgo collaboration has estimated its component masses to be $5.7^{+4.0}_{-2.3}\,\msun$ and $1.5^{+0.8}_{-0.5}\,\msun$, rendering it the first candidate BH-NS merger. The data are uninformative about potential tidal effects, parametrized by \citep{FlanaganHinderer2008}
\begin{equation}
\tilde{\Lambda}=\frac{32}{39}\frac{M_{\rm NS}^4(M_{\rm NS}+12M_{\rm BH})}{(M_{\rm NS}+M_{\rm BH})^5}\frac{k_2}{C_{\rm NS}}\,,
\label{eqn:gwtidaleff}
\end{equation}
where $k_2$ is the second Love number. The effective spin of the systems has been measured to $-0.03^{+0.33}_{-0.30}$. 

\begin{figure} 
\centering
\includegraphics[scale=0.565]{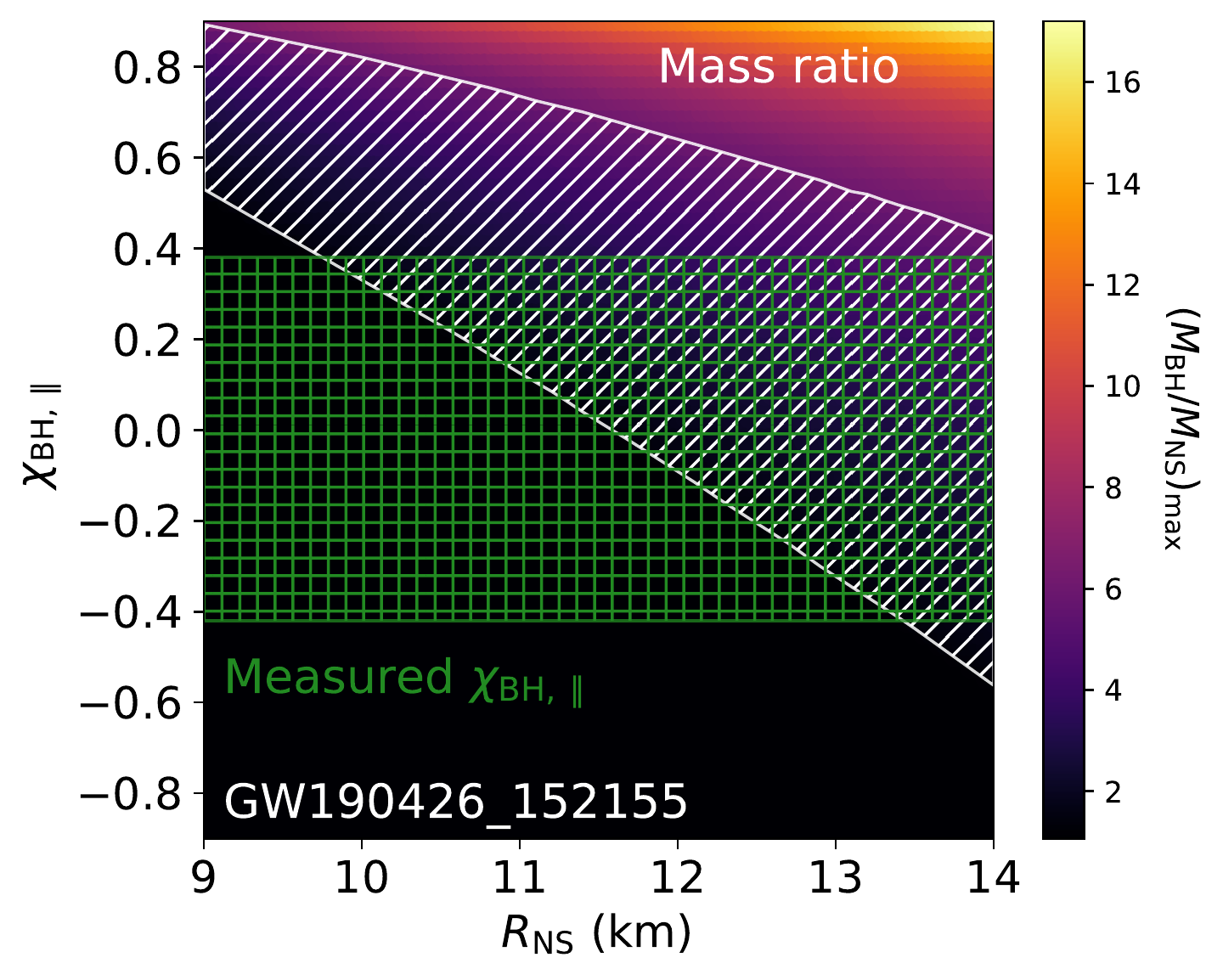}
\caption{The case of the candidate BH-NS merger GW190426\_152155 from GWTC-2 \citep{AbbottAbbott2020a}, for which no EM counterpart has been observed \citep{AbbottAbbott2020d}. Measurement errors on the effective spin and mass ratio are too large to place an upper limit on the NS radius for this candidate event.}
\label{fig:plot3}
\end{figure}

Figure~\ref{fig:plot3} illustrates the case our method is applied to GW190426\_152155 \citep{AbbottAbbott2020a}. We both show $\chi_{\rm BH,\parallel}$, derived from the measured effective spin $-0.03^{+0.33}_{-0.30}$, and the BH-to-NS mass ratio $3.35^{+2.54}_{-2.22}$, as derived from LIGO/Virgo analysis. Note that effective spin and mass ratio are strongly correlated. For this event, no EM counterpart has been observed \citep{AbbottAbbott2020d}, implying a plunging of the NS onto the BH. Measurement errors on the effective spin and mass ratio are too large to place an upper limit on the NS radius for this candidate event. Therefore, GW190426\_152155 is not a good candidate to get information on the NS equation of state from current observations. If an EM counterpart had been observed, we would have been able to place a lower limit on the NS radius, $R_{\rm NS}\gtrsim 10$\,km.

\section{Conclusions}
\label{sect:concl}

BH-NS mergers are interesting since they could produce an EM counterpart in the form of a short gamma-ray burst (GRB), which can provide crucial information on their origin and NS structure. Despite the growing number of events, there is only one candidate BH-NS merger, namely GW190426\_152155.

We have shown how to use information on the EM counterpart in BH-NS mergers can be used to place constraints on the NS radius. In particular, we have illustrated how to derive a lower (upper) limit based on the presence (absence) of an accretion disk that powers the EM counterpart \citep{FoucartHinderer2018}. We have concluded that our method works best when the aligned component of the BH spin is the largest. \citet{AscenziDeLillo2019} presented a method based on a similar idea in \citet{PannaraleOhme2014} to exploit joint GW-short GRB observations of BH-NS coalescences. \citet{HindererNissanke2019} performed a similar analysis on GW170817, working under the assumption that
the event originated from a BH-NS coalescence, and exploited
the EM constraints from the kilonova light curve. We have also applied our method to the case of the candidate LIGO/Virgo event GW190426\_152155, for which no EM counterpart has been observed \citep{AbbottAbbott2020d}. We have found that large measurement errors on the spin and mass ratio prevent from placing an upper limit on the NS radius for this candidate event. \citet{LiHan2020} reanalyzed GW190426\_152155 using several waveforms with different characteristics and found that the results are influenced by the priors of mass ratio, showing that the chance of observing an EM counterpart is rather unpromising for GW190426\_152155. \citet{CoughlinDietrich2020} employed three different kilonova models and derived upper bounds on the ejecta mass for some of the LIGO candidate events, which they used to put constraints on the mass-ratio, spin, and NS compactness for GW190426\_152155 \citep[for recent candidates for more recent candidates see][]{AnandCoughlin2021}.

We finally note that we have assumed an EM counterpart will be seen in the case of tidal disruption. However, this might not be the case due to observational strategies and limitations. The current search methods in the LIGO/Virgo pipeline include both modelled searches for a GW signal (compatible with the inspiral of a NS-NS or BH-NS binary) within $6$\,s of data associated with an observed short GRB and unmodelled search for generic transients, consistent with the sky localization and time window for each GRB, that begins $600$\,s before the GRB trigger time and ends $60$\,s after it \citep[see][and references therein]{AbbottAbbott2020d}. A GRB not pointed at us would not be detectable. Disrupting BH-NS binaries could also have a lot of dynamical ejecta, which may power a kilonova. However, follow-up observations of potential BH-NS mergers are so far only covering a small portion of the sky consistent with the detected GW signals and are typically not deep enough to observe all kilonovae. Thus, the lack of an observed EM signal following a BH-NS merger event is significantly less constraining in our model. Finally, existing theoretical models remain limited by the lack of understanding of post-merger outflows and by nuclear physics and radiation transport uncertainties \citep[e.g.,][]{BarnesKasen2016,BarbieriSalafia2020}.

While obtaining precise models for the observable EM signals powered by BH-NS binaries can be challenging, the dependence of these signals on the properties of BH-NS binaries is essential for extracting valuable information \citep[e.g.,][]{AscenziDeLillo2019,HindererNissanke2019,SridharZrake2021}. This can complement the information that can be obtained from the GW signal, either from the potential tidal dephrasing or the cut-off frequency when disruption occurs ($\sim 1$\,kHz), to help constrain the equation of state of NSs \citep{LackeyKyutoku2014,PannaraleBerti2015}.

\section*{Acknowledgements}
%\acknowledgements

GF acknowledges support from CIERA at Northwestern University. This work was supported in part by Harvard's Black Hole Initiative, which is funded by grants from JFT and GBMF.\\
\ \\
\textit{Data Availability}\\
\ \\
The data underlying this article will be shared on reasonable request to the corresponding author.\\

\bibliographystyle{mn2e}
\bibliography{refs}

\end{document}